\documentclass[12pt]{article}
\usepackage[utf8]{inputenc}
\usepackage{amsmath,amsfonts,amssymb,amsthm}
\usepackage{mathrsfs}
\usepackage{booktabs}
\usepackage{empheq}
\usepackage{appendix}
\usepackage{csquotes}
\usepackage{titlesec}
\usepackage{hyperref}
\usepackage{comment}
\usepackage{setspace}
\usepackage{bbm}
\usepackage{bm}
\usepackage{tikz}
\usepackage{pgfplots}
\hypersetup{
     colorlinks=true,
     linkcolor=black,
     filecolor=blue,
     citecolor = blue,      
     urlcolor=blue,
     }

\usepackage[nameinlink]{cleveref}
\newcommand\blfootnote[1]{%
  \begingroup
  \renewcommand\thefootnote{}\footnote{#1}%
  \addtocounter{footnote}{-1}%
  \endgroup
}
\newtheorem{theorem}{Theorem}[section]

\newtheorem{assumption}{Assumption}[section]

\newtheorem{lemma}{Lemma}[section]

\definecolor{cyan}{cmyk}{1, 0.4, 0, 0}

\usepackage[uniquename=false,minnames=1,maxnames=2,backend=biber,style=authoryear,natbib=true]{biblatex}

\usepackage{geometry}
\emergencystretch=\maxdimen
\def\Perp{\perp\!\!\!\perp}

\title{Nonparametric Testability of Slutsky Symmetry \blfootnote{A previous version was titled ``A nonparametric test of Slutsky Symmetry''. The authors thank Shomu Banerjee, Stefan Hoderlein, and audiences at Emory, LMU Munich, NYU, the University of Michigan, the 4th Georgia Econometrics workshop, the Econometrics mini conference at the University of Iowa, the 2025 IAAE Annual Conference in Turin, Italy, the 2025 World Congress of the Econometric Society and the 2025 Canadian Econometric Study Group Meeting for helpful feedback. All errors are the authors'.}}
\author{Florian Gunsilius \\Department of Economics\\Emory University\\ \href{mailto:fgunsil@emory.edu}{\texttt{fgunsil@emory.edu}}
\and
Lonjezo Sithole \\Department of Economics\\University of Michigan\\ \href{mailto:lsithole@umich.edu}{\texttt{lsithole@umich.edu}}}

\begin{document}

\maketitle
\begin{abstract}
   Economic theory implies strong limitations on what types of consumption
behavior are considered rational. Rationality implies that the Slutsky matrix, which
captures the substitution effects of compensated price changes on demand for different
goods, is symmetric and negative semi-definite. While empirically informed versions
of negative semi-definiteness have been shown to be nonparametrically testable, the
analogous question for Slutsky symmetry has remained open. Recently, it has even
been shown that the symmetry condition is not testable via the average Slutsky
matrix, prompting conjectures about its non-testability. We settle this question by
deriving nonparametric conditional quantile restrictions on observable data that
constitute a testable implication of Slutsky symmetry in an
empirical setting with individual heterogeneity and endogeneity. The theoretical
contribution is a multivariate generalization of identification results for partial effects
in nonseparable models without monotonicity, which is of independent interest. This
result has implications for different areas in econometric theory, including
nonparametric welfare analysis with individual heterogeneity for which, in the case of
more than two goods, the symmetry condition introduces nonlinear correction factors.

\vspace{0.1in}

\end{abstract}

\newpage

\section{Introduction}

Rationality is the foundational assumption of consumer demand theory. A smooth demand function is rationalizable, that is, consistent with utility maximization, if and only if the associated Slutsky matrix of compensated price effects is symmetric and negative semidefinite \citep{samuelson1948consumption, houthakker1950revealed, hurwicz1971integrability}. Whether these conditions hold in data is therefore a question of first-order importance for applied welfare analysis.

Testing the Slutsky conditions nonparametrically, i.e., without committing to a functional form for demand, has proven difficult. For negative semidefiniteness, \citet{DETTE2016129} provide a nonparametric test in an empirical model with individual heterogeneity and endogeneity. For symmetry, no analogous result has been available to date. Parametric tests exist \citep{barten1967evidence, laitinen1978demand, MEISNER1979231, BERA1981101, bewley1983tests, TAYLOR1985327, silver1989testing}, but conflate violations of symmetry with misspecification. Nonparametric approaches such as \citet{lewbel1995consistent}, \citet{HAAG200933}, and \citet{hoderlein2011many} rely on additional restrictions that are either unverifiable or impose substantial structure on the demand system. Recently, \citet{kono2025untestability} shows that symmetry cannot be tested from the average Slutsky matrix alone, and \citet{maes2024beyond} conjecture that in the general setting of \citet{DETTE2016129}, nonparametric testability of Slutsky symmetry is impossible.

We show that this conjecture is false in the general setting. We derive a nonparametric restriction on observable conditional and marginal quantiles that is necessary for Slutsky symmetry, providing a testable characterization of the symmetry condition in the empirical setting of \citet{DETTE2016129}. The key theoretical contribution is a bivariate extension of the identification result of \citet{Hoderlein2007} for average partial effects in nonseparable models without monotonicity. This extension, which is of independent interest, reveals that in the multivariate setting the identification result acquires correction terms that are absent in the univariate case. These terms have concrete economic content: they capture how the composition of preference types at a given demand level shifts with prices, and they govern the gap between conditional quantile demands and true Hicksian demands in multi-good welfare analysis.

Our approach is based on continuous demands in a nonseparable model with unrestricted heterogeneity. \citet{hausman2016individual} provide the foundational framework for nonparametric welfare analysis in this setting, but their analysis is restricted to two goods where symmetry holds by construction. We generalize their approach to multiple goods, where symmetry has empirical content and must be verified. 

\section{A nonparametric Slutsky symmetry condition}
This section contains the main identification result for the Slutsky symmetry condition, which leads to a testable nonparametric restriction. To state it, we start with the required assumptions on the data and an extension of the ideas in \citet{Hoderlein2007} to conditional quantiles. 

\subsection{Setup and assumptions}

The key idea is based on the connection between quantile functions of the data-generating process and properties of the Slutsky matrix in an empirical model with individual heterogeneity and endogeneity. \citet{DETTE2016129} posit an IV model for a nonparametric demand system with a potentially endogenous covariate (for example, income or total expenditures).  The endogeneity of total expenditure is, in part, because categories of individual expenditures are often mismeasured, and hence the aggregate expenditure measure is mismeasured. Moreover, the authors model expenditure while allowing for unobserved individual heterogeneity. In the sequel, we carry over the notation and model from \citet{DETTE2016129} to offer a unified treatment:

\begin{equation}\label{eq:model}
\begin{aligned}
&Y=\psi(P, X, U)=\psi(P, X, \nu(Q, A))=\phi(P, X, Q, A) \\
&X=\mu(P, Q, S, V).
\end{aligned}
\end{equation}
Here $Y$, a vector of quantities of goods demanded, takes values in $\mathbb{R}^{L-1}$; and $X$ is an endogenous variable, for instance income. We assume that there exists exactly one exogenous shifter $S$ for the endogenous variable. This is done for ease of exposition, analogous to the result in \citet{DETTE2016129}, and can be generalized to higher dimensions \citep[e.g.][]{gunsilius2023condition}. $V$ captures the unobservable variation in this first stage regression. $V$ can be solved for and used as residuals in a control function approach. For standard identification arguments \citep[e.g.][]{imbens2009identification}, therefore, we assume $\mu$ is invertible in its last argument. $P \in \mathbb{R}^{L-1}$ is the vector of prices for the goods; and $U$ is an unrestricted unobservable describing the preferences. Making $U$ as general as possible is crucial to account for any form of preference function. By writing $U=\nu(Q,A)$, we assume that the preference function is a function of observable characteristics $Q$, like age for instance, and unobservable characteristics $A$. $A$ can therefore be interpreted as residual unobserved heterogeneity after controlling for the observable part of individual heterogeneity. 
Putting everything together, we have the same assumption on the model as \citet{DETTE2016129}, to which we refer for further detail. 

\begin{assumption}[Model]\label{ass:mod}
    Let $(\Omega, \mathcal{F}, P)$ be a complete probability space on which we define the random variables $A : \Omega \to \mathcal{A}\subseteq \mathbb{R}^\infty$, and $(Y, P, X, Q, S, V) : \Omega \to \mathcal{Y} \times \mathcal{P} \times \mathcal{X} \times \mathcal{Q} \times \mathcal{S} \times \mathcal{V}, \mathcal{Y} \subseteq \mathbb{R}^{L-1}, \mathcal{P} \subseteq \mathbb{R}^{L-1}, \mathcal{X} \subseteq \mathbb{R}, \mathcal{Q} \subseteq \mathbb{R}^K, \mathcal{S} \subseteq \mathbb{R}, \mathcal{V} \subseteq \mathbb{R},$ with $L$ and $K$ finite integers, such that
\begin{align*}
Y &= \phi(P, X, Q, A)\\
X &= \mu(P, Q, S, V),
\end{align*}
where $\phi : \mathcal{P} \times \mathcal{X} \times \mathcal{Q} \times \mathcal{A} \to \mathcal{Y}$ and $\mu : \mathcal{P} \times \mathcal{Q} \times \mathcal{S} \times \mathcal{V} \to \mathcal{X}$ are Borel functions, and realizations of $(Y, P, X, Q, S)$ are observable, whereas $(A, V)$ are latent variables. The function $\mu$ is invertible in its last argument, for every $(p, q, s)$.
\end{assumption}

The Slutsky matrix $\mathcal{S}(p, x, u)$ takes the form
\[
\mathcal{S}=D_{p} \psi(p, x, u)+\partial_{x} \psi(p, x, u) \psi(p, x, u),
\]
where $D_{p}$ is the Jacobian of Marshallian demands with respect to prices and $\partial_x$ is a vector of partial derivatives of Marshallian demands with respect to income.\\
Following \citet{DETTE2016129}, we denote $W=(P, X, Q, V)$ and introduce $k(\alpha, b \mid w)$ as the $\alpha$-quantile defined as

\[
P\left(b^{\prime} Y \leq k(\alpha, b \mid w) \mid W=w\right)=\alpha\]
for any nonzero $b \in \mathbb{R}^{L-1}$. 
To derive a condition on the observable quantiles that captures the symmetry condition of the Slutsky matrix, only considering univariate marginal quantiles is not enough, as one has to compare off-diagonal terms. However, we can focus on only two elements at a time to capture the symmetry condition. Still, only considering bivariate joint quantiles does not provide identification of the relevant conditions, because the result in \citet{Hoderlein2007} cannot be generalized to the multivariate setting directly \citep{Hoderlein2009}. 

We show that symmetry becomes testable in this setting by considering a combination of \emph{conditional- and marginal quantiles}, for which we provide a generalization of the result in \citet{Hoderlein2007} in section \ref{sec:hod_mam}.
For symmetry, we only need to consider the case where $b$ is a unit vector, that is, we only need to consider elements of the form $e_{i}^{\prime} Y$ for $e_{i} \in \mathbb{R}^{L-1}$ the $i$-th unit vector in $\mathbb{R}^{L-1}$. In this case the above defined $\alpha$-quantile reduces to the $\alpha$-quantile of the marginal distribution of $Y_{i}=e_{i}^{\prime} Y$. 
For notational purposes, we define the \emph{marginal $\alpha$-quantile} as
     
\begin{equation*}
P\left(Y_{i}  \leq k_{\alpha, i}(w) \mid W=w\right)=\alpha 
\end{equation*}

In the same vein, we define the \emph{conditional $\gamma$-quantile} of the element $Y_{i} \equiv e_{i}^{\prime} Y$ given that the element $Y_{j} \equiv e_{j}^{\prime} Y$ lies at the $\alpha$ marginal level by

\begin{equation*}
P\left(Y_{i} \leq k_{\gamma, i}\left(w, k_{\alpha, j}(w)\right) \mid W=w, Y_{j}=k_{\alpha, j}(w)\right)=\gamma. 
\end{equation*}

This conditional quantile depends not only on $W$ but also on the marginal quantile $k_{\alpha, j}(w)$, which in turn also depends on $W$. In fact, it will be the interplay of the conditional and the marginal quantile which will enable us to derive conditions on the observable data that captures the Slutsky symmetry. 
We also need to make use of the \emph{joint bivariate $\alpha$-quantile} $K_{i j}(\alpha \mid w)$ for elements $\left(Y_{i}, Y_{j}\right) \equiv\left(e_{i}^{\prime} Y, e_{j}^{\prime} Y\right)$ of the vector $Y \in \mathbb{R}^{L-1}$, which is defined as
\begin{equation*}
P\left(\left(Y_{i}, Y_{j}\right) \leq K_{i j}(\alpha \mid w) \mid W=w\right)=\alpha. 
\end{equation*}
Note that $K_{i j}$ is not unique without further assumption: it is not a point but an isoquant in $\mathbb{R}^{2}$, which is the main complication for testing symmetry nonparametrically.

Since we generalize the identification result in \citet{Hoderlein2007} to conditional and marginal quantiles, we aim to keep the same notation and assumptions. These assumptions are also stated in the same notation as the ones in \citet{DETTE2016129} to keep the two results unified. We set $Z:=(Q, V)$. 

\begin{assumption}[Independence]\label{ass:ind} 
 $A \Perp(P, X) \mid Z$.   
\end{assumption}
The independence assumption requires that after controlling for observable preference characteristics and unobservable first-stage, the residual unobserved heterogeneity represented by $A$ is independent of prices and the endogenous variable. This conditional independence assumption is key to identification; it maps the space of nonseparable demand functions, which is inherently unobservable, to something we can observe: regression quantiles. It is this  feature that allows us to derive a testable restriction on the quantiles of the data generating process, develop a test based on the  restriction and infer Slutsky symmetry for the unobserved nonseparable demand functions. 

In the following, we denote by $\partial_{w_{1} s} k_{\alpha, j}(w^*)$ the partial derivative of the quantile $k_{\alpha, j}$ with respect to the price of good s (the $s^{th}$ element in the price vector, which in turn is the first element of $w$) at $w^{*}$. For the conditional quantile $k_{\gamma, i}\left(w^{*}, k_{\alpha, j}\left(w^{*}\right)\right)$, we define by $\partial_{1, w_{1 s}} k_{\gamma, i}\left(w^{*}, k_{\alpha, j}\left(w^{*}\right)\right)$ the partial derivative with respect to price of good $s$ (the $s^{th}$  element of the price vector, which in turn is the first argument of $w$) at $w^{*}$ and by $\partial_{2} k_{\gamma, i}\left(w^{*}, k_{\alpha, j}\left(w^{*}\right)\right)$ the partial derivative with respect to the second argument, i.e. $k_{\alpha, j}\left(w^{*}\right)$. 

Furthermore, we write the conditional density $f_{e_{i}^{\prime} Y \mid W, e_{j}^{\prime} Y}\left(y_{i}\right)$ at a point $\left(w^{*}, k_{\alpha, j}\left(w^{*}\right)\right)$ as $f_{e_{i}^{\prime} Y \mid W=w^{*}, e_{j}^{\prime} Y=k_{\alpha, j}\left(w^{*}\right)}\left(y_{i}\right)$. Similarly, for the conditional distribution function $F_{Y_i \mid W, Y_j = y_j}(y_i)$, we write $\nabla_{p,1} F_{Y_i \mid W=w^*, Y_j = y_j}(y_i)$ for the gradient with respect to the price components of $W$ holding the conditioning value $y_j$ fixed, $\partial_{x,1} F_{Y_i \mid W=w^*, Y_j = y_j}(y_i)$ for the partial derivative with respect to income holding $y_j$ fixed, and $\partial_2 F_{Y_i \mid W=w^*, Y_j = y_j}(y_i)$ for the partial derivative with respect to the conditioning value $y_j$. We impose regularity conditions on the conditional distribution function, the conditional density, the conditional quantile and the demand function, which we have relegated to the appendix. 

\subsection{A generalization of \citet{Hoderlein2007}}\label{sec:hod_mam}
We can now state a multivariate extension of the identification result in \citet{Hoderlein2007}:

\begin{lemma}\label{lem:hod_mam}
Under Assumption~\ref{ass:mod} and Assumption~\ref{ass:ind} along with the regularity conditions in Assumption~\ref{ass:reg}, fix a point $(y_i^*, y_j^*)$ in the interior of the support of $(Y_i, Y_j)$ given $W = w^*$ and define
\[
\alpha_j := F_{Y_j|W=w^*}(y_j^*), \qquad \gamma_{i|j} := F_{Y_i|W=w^*, Y_j = y_j^*}(y_i^*),
\]
so that $k_{\alpha_j,j}(w^*) = y_j^*$ and $k_{\gamma_{i|j},i}(w^*, k_{\alpha_j,j}(w^*)) = y_i^*$. Then for $L \geq 3$, the following holds
\begin{align*}
&E\!\left[\partial_{w_{1s}}\phi \;\middle|\; W = w^*, \, e_i'Y = k_{\gamma_{i|j},i}(w^*, k_{\alpha_j,j}(w^*)), \, e_j'Y = k_{\alpha_j,j}(w^*)\right] \\
&= \partial_{1,w_{1s}} k_{\gamma_{i|j},i}(w^*, k_{\alpha_j,j}(w^*)) \\
&\quad + \partial_{w_{1s}} k_{\alpha_j,j}(w^*) \left[\partial_2 k_{\gamma_{i|j},i}(w^*, k_{\alpha_j,j}(w^*)) + \frac{\partial_2 F_{Y_i|W=w^*, Y_j = k_{\alpha_j,j}(w^*)}(k_{\gamma_{i|j},i}(w^*, k_{\alpha_j,j}(w^*)))}{f_{Y_i|W=w^*, Y_j = k_{\alpha_j,j}(w^*)}(k_{\gamma_{i|j},i}(w^*, k_{\alpha_j,j}(w^*)))}\right] \\
&\quad + \frac{\partial_{w_{1s}} F_{Y_i|W=w^*, Y_j = k_{\alpha_j,j}(w^*)}(k_{\gamma_{i|j},i}(w^*, k_{\alpha_j,j}(w^*)))}{f_{Y_i|W=w^*, Y_j = k_{\alpha_j,j}(w^*)}(k_{\gamma_{i|j},i}(w^*, k_{\alpha_j,j}(w^*)))}
\end{align*}
for $s = 1, \ldots, L$.
\end{lemma}

This result is analogous to the univariate version of \citet{Hoderlein2007}. The difference is that the conditional quantile is affected through several paths when changing $w$, instead of just the marginal $k_{\alpha, j}\left(w\right)$. This results in the additional complex correction terms in the expression. The underlying idea of the proof lies in the fact that the inherently $2$-dimensional quantile function $K_{ij}(\alpha, w)$ can be represented in two different ways by the two respective marginal- and the corresponding conditional quantiles. This representation is unique for regular quantiles. 

Lemma \ref{lem:hod_mam} can be straightforwardly extended to more than two dimensions by repeated conditioning. In this case, the expression becomes even more complicated with additional terms; this stems from the fact that for $d$-dimensions, one needs to iteratively condition on $d-1$ quantiles. Since we only care about the Slutsky symmetry case, which is bivariate, we omit the d-dimensional version. Note that this result does not hinge on the generality of Assumption \ref{ass:mod}: in applications without endogeneity, this result still applies with a \enquote{reduced form} nonparametric demand model (the second stage model in Assumption \ref{ass:mod}).

\subsection{A Slutsky symmetry condition on conditional and marginal quantiles}

Lemma \ref{lem:hod_mam} now enables us to derive the condition on the quantile functions which captures the symmetry condition of the Slutsky matrix. Based on Lemma \ref{lem:hod_mam}, the intuition is as follows: the joint quantile $K_{ij}(\alpha|w)$ can be represented uniquely in two ways, corresponding to two different combinations of marginal- and conditional quantiles, which captures the ``off-diagonal'' terms in the Slutsky matrix via these marginal and corresponding conditional quantile functions. This leads to the following representation:

 \begin{theorem} [Slutsky symmetry]\label{thm:symmetry}
 Let Assumptions~\ref{ass:mod}--\ref{ass:reg} hold. Fix a point $(y_i^*, y_j^*)$ in the interior of the support of $(Y_i, Y_j)$ given $W = w^*$, and define the following quantile indices:
\begin{align*}
\alpha_i &:= F_{Y_i|W=w^*}(y_i^*), \qquad \alpha_j := F_{Y_j|W=w^*}(y_j^*), \\
\gamma_{i|j} &:= F_{Y_i|W=w^*, Y_j = y_j^*}(y_i^*), \qquad \gamma_{j|i} := F_{Y_j|W=w^*, Y_i = y_i^*}(y_j^*),
\end{align*}
so that $k_{\alpha_i, i}(w^*) = y_i^*$, $k_{\alpha_j, j}(w^*) = y_j^*$, $k_{\gamma_{i|j}, i}(w^*, k_{\alpha_j, j}(w^*)) = y_i^*$, and $k_{\gamma_{j|i}, j}(w^*, k_{\alpha_i, i}(w^*)) = y_j^*$.

Define the correction terms
\begin{align*}
C_{ij}(w^*) &:= \partial_2 k_{\gamma_{i|j}, i}(w^*, k_{\alpha_j, j}(w^*)) + \frac{\partial_2 F_{Y_i|W=w^*, Y_j = k_{\alpha_j,j}(w^*)}(k_{\gamma_{i|j},i}(w^*, k_{\alpha_j,j}(w^*)))}{f_{Y_i|W=w^*, Y_j = k_{\alpha_j,j}(w^*)}(k_{\gamma_{i|j},i}(w^*, k_{\alpha_j,j}(w^*)))}, \\[6pt]
C_{ji}(w^*) &:= \partial_2 k_{\gamma_{j|i}, j}(w^*, k_{\alpha_i, i}(w^*)) + \frac{\partial_2 F_{Y_j|W=w^*, Y_i = k_{\alpha_i,i}(w^*)}(k_{\gamma_{j|i},j}(w^*, k_{\alpha_i,i}(w^*)))}{f_{Y_j|W=w^*, Y_i = k_{\alpha_i,i}(w^*)}(k_{\gamma_{j|i},j}(w^*, k_{\alpha_i,i}(w^*)))}, \\[6pt]
D_{ij}(w^*) &:= \frac{\nabla_{p,1} F_{Y_i|W=w^*, Y_j = k_{\alpha_j,j}(w^*)}(k_{\gamma_{i|j},i}(w^*, k_{\alpha_j,j}(w^*)))}{f_{Y_i|W=w^*, Y_j = k_{\alpha_j,j}(w^*)}(k_{\gamma_{i|j},i}(w^*, k_{\alpha_j,j}(w^*)))}, \\[6pt]
D_{ij}^{(x)}(w^*) &:= \frac{\partial_{x,1} F_{Y_i|W=w^*, Y_j = k_{\alpha_j,j}(w^*)}(k_{\gamma_{i|j},i}(w^*, k_{\alpha_j,j}(w^*)))}{f_{Y_i|W=w^*, Y_j = k_{\alpha_j,j}(w^*)}(k_{\gamma_{i|j},i}(w^*, k_{\alpha_j,j}(w^*)))}, \\[6pt]
D_{ji}(w^*) &:= \frac{\nabla_{p,1} F_{Y_j|W=w^*, Y_i = k_{\alpha_i,i}(w^*)}(k_{\gamma_{j|i},j}(w^*, k_{\alpha_i,i}(w^*)))}{f_{Y_j|W=w^*, Y_i = k_{\alpha_i,i}(w^*)}(k_{\gamma_{j|i},j}(w^*, k_{\alpha_i,i}(w^*)))}, \\[6pt]
D_{ji}^{(x)}(w^*) &:= \frac{\partial_{x,1} F_{Y_j|W=w^*, Y_i = k_{\alpha_i,i}(w^*)}(k_{\gamma_{j|i},j}(w^*, k_{\alpha_i,i}(w^*)))}{f_{Y_j|W=w^*, Y_i = k_{\alpha_i,i}(w^*)}(k_{\gamma_{j|i},j}(w^*, k_{\alpha_i,i}(w^*)))}.
\end{align*}

If $\mathcal{S}$ is symmetric, then the following equality holds for all $i, j = 1, \ldots, L-1$ and all $(y_i^*, y_j^*)$ in the interior of the support:
\begin{align}
&\nabla_{p,1} k_{\gamma_{i|j},i}(w^*, k_{\alpha_j,j}(w^*)) \, e_j + \partial_{x,1} k_{\gamma_{i|j},i}(w^*, k_{\alpha_j,j}(w^*)) \, k_{\alpha_j,j}(w^*) \notag \\
&\quad + C_{ij}(w^*) \cdot \Big[\nabla_p k_{\alpha_j,j}(w^*) \, e_j + \partial_x k_{\alpha_j,j}(w^*) \, k_{\alpha_j,j}(w^*)\Big] \notag \\
&\quad + D_{ij}(w^*) \cdot e_j + D_{ij}^{(x)}(w^*) \, k_{\alpha_j,j}(w^*) \notag \\[6pt]
&= \nabla_{p,1} k_{\gamma_{j|i},j}(w^*, k_{\alpha_i,i}(w^*)) \, e_i + \partial_{x,1} k_{\gamma_{j|i},j}(w^*, k_{\alpha_i,i}(w^*)) \, k_{\alpha_i,i}(w^*) \notag \\
&\quad + C_{ji}(w^*) \cdot \Big[\nabla_p k_{\alpha_i,i}(w^*) \, e_i + \partial_x k_{\alpha_i,i}(w^*) \, k_{\alpha_i,i}(w^*)\Big] \notag \\
&\quad + D_{ji}(w^*) \cdot e_i + D_{ji}^{(x)}(w^*) \, k_{\alpha_i,i}(w^*). \label{eq:symmetry_condition}
\end{align}
\end{theorem}

 Equation \eqref{eq:symmetry_condition} gives a necessary condition for Slutsky Symmetry in the empirical setting described by Model \eqref{eq:model} and Assumptions \ref{ass:reg} and \ref{ass:ind}. If this condition is not satisfied for any $i,j$ pair of goods under the model assumptions, then the Slutsky symmetry is not satisfied for that set of goods in this model. The complication arises from the correction terms $C_{ij}$, $C_{ji}$, $D_{ij}$, $D_{ji}$, $D_{ij}^{(x)}$, and $D_{ji}^{(x)}$, which do not appear in the negative semidefinite case.

\section{Slutsky Symmetry and Nonparametric Welfare Analysis}

Our results yield new insights for nonparametric welfare analysis with individual heterogeneity in multi-good settings. \citet{hausman2016individual} show that in a two-good setting with a numeraire, quantile demands coincide with true demands, using the univariate identification result of \citet{Hoderlein2007} and the negative semidefiniteness condition of \citet{DETTE2016129}. In that setting, Slutsky symmetry holds by construction. With more than two goods, symmetry has empirical content, and the univariate identification argument no longer suffices. Existing approaches either assume symmetry outright or impose restrictive conditions such as monotonicity in scalar unobserved heterogeneity.

Our bivariate extension of \citet{Hoderlein2007} in Lemma~\ref{lem:hod_mam}, together with the symmetry condition in~\eqref{eq:symmetry_condition}, restores the quantile-based approach to welfare analysis under general heterogeneity. Theorem~\ref{thm:symmetry} shows that with vector-valued demands, true demands are identified by conditional and marginal quantiles satisfying~\eqref{eq:symmetry_condition}. Unlike in \citet{hausman2016individual}, where marginal quantile demands coincide with true demands directly, conditional quantile demands for each good coincide with true demands only up to the correction factors $C_{ij}(w^*)$, $D_{ij}(w^*)$, and $D_{ij}^{(x)}(w^*)$ induced by conditioning on the demand level of the other good.

\paragraph{Empirical content of the condition} The symmetry condition in Theorem~\ref{thm:symmetry} involves two types of correction terms beyond the leading conditional and marginal quantile derivatives: the terms $C_{ij}(w^*)$ and $D_{ij}(w^*)$ (along with their income counterparts $D_{ij}^{(x)}(w^*)$). These terms have distinct economic content that is worth unpacking, as they govern the gap between conditional quantile demands and true Hicksian demands in multi-good settings.

\paragraph{The $C_{ij}$ terms.} The term $C_{ij}(w^*)$ captures how the conditional quantile of $Y_i$ responds to shifts in the level of $Y_j$ at which we condition. It combines a direct effect ($\partial_2 k_{\gamma_{i|j},i}$, the mechanical response of the conditional quantile to a change in the conditioning value) with an indirect effect through the conditional distribution. Economically, $C_{ij}(w^*)$ is small when the demand for good $i$ at a given quantile is approximately insensitive to the demand level of good $j$. This is plausible when the two goods are weakly related in preferences (for instance, under approximate weak separability) or when each good individually constitutes a small fraction of total expenditure, so that the income reallocation channel linking their demands is muted.

\paragraph{The $D_{ij}$ terms.} The terms $D_{ij}(w^*)$ and $D_{ij}^{(x)}(w^*)$ capture a fundamentally different phenomenon: how the conditional distribution of $Y_i$ given $Y_j = y_j^*$ shifts when prices or income change, \emph{holding the conditioning level $y_j^*$ fixed}. This effect is absent in the univariate setting of \citet{Hoderlein2007}, where the independence assumption $A \perp\!\!\!\perp (P,X) \mid Z$ suffices to eliminate any such dependence. In the bivariate setting, the conditional distribution $F_{Y_i | W, Y_j = y_j^*}$ conditions on the level set $\{a : e_j'\phi(p, x, z^*, a) = y_j^*\}$, which shifts with $(p,x)$ through the demand function $\phi$, even though the marginal distribution of the preference heterogeneity $A$ given $Z$ does not depend on prices or income.

Economically, $D_{ij}(w^*) = 0$ when the \emph{composition of preference types} demanding a fixed quantity $y_j^*$ of good $j$ is stable with respect to price changes. This is a natural condition in settings where the mapping from preferences to demand for good $j$ is approximately one-to-one (so that the level set $\{a : e_j'\phi(\cdot, a) = y_j^*\}$ shifts rigidly rather than changing its composition), or more generally when the heterogeneity in demand for good $i$ among consumers who demand the same quantity of good $j$ is not systematically related to the price sensitivity of demand for good $j$. A sufficient condition for $D_{ij}(w^*) = 0$ for all $w^*$ is that $\phi$ is monotone in the component of $A$ that drives $Y_j$, reducing the bivariate problem to the univariate case of \citet{Hoderlein2007}. The results of this paper, when integrated with existing negative-semidefiniteness tests, now allow researchers to comprehensively verify Slutsky conditions prior to welfare analysis with unobserved heterogeneity.

\section{Conclusion}
We provide a fully nonparametric analysis of the testability of the symmetry of the Slutsky matrix via empirical quantiles, complementing existing work that provides a test for the negative semidefiniteness. This work, therefore, closes the gap on testing an empirically informed version of rationality,  providing a testable symmetry condition relevant for nonparametric welfare analysis with individual heterogeneity. Incidentally, the results show that the symmetry condition is testable using \emph{conditional} quantile restrictions. This shows that a recent conjecture \citep{maes2024beyond} is  false and opens the door to testing a empirically informed version of rationality with heterogeneity. The symmetry condition provides empirical restrictions in empirical welfare analysis. Since it involves only conditional and marginal quantiles and their derivatives, all of its components are estimable from data via conditional quantile inference methods. In the multi-good case, the condition provides nonlinear correction factors that govern the gap between conditional quantile demands and true Hicksian demands, which must be accounted for in any nonparametric welfare analysis with individual heterogeneity beyond the two-good setting.

\printbibliography

\appendix

\section{Proofs}
\subsection{Regularity Assumptions}

\begin{assumption} [Regularity]\label{ass:reg}
    For fixed $w^{*}:=\left(p^{*}, x^{*}, z^{*}\right) \in \mathcal{P} \times \mathcal{X} \times \mathcal{Z}$ all $i, j=1,2, \ldots, L-1$, and $\gamma, \alpha \in[0,1]$ the following conditions hold:

\begin{enumerate}
  \item The conditional distribution function $F_{Y_{i} \mid W, Y_{j}}$ of $e_{i}^{\prime} Y$ given $\left(W, e_{j}^{\prime} Y\right)$ is absolutely continuous with respect to Lebesgue measure for $\left(p, x, e_{j}^{\prime} y\right)$ in a neighborhood of $\left(w^{*}, k_{\alpha, j}\left(w^{*}\right)\right)$  and for $z=z^{*}$.
  \item The conditional density $f_{e_{i}^{\prime} Y \mid W, e_{j}^{\prime} Y}\left(y_{i}\right)$ of $e_{i}^{\prime} Y$ given $\left(W, e_{j}^{\prime} Y\right)$ is continuous in $\left(e_{i}^{\prime} y\right)$ and differentiable in $\left(P, X, e_{j}^{\prime} Y\right)$ in a neighborhood of the point $\left(k_{\gamma, i}\left(w^{*}, k_{\alpha, j}\left(w^{*}\right)\right), k_{\alpha, j}\left(w^{*}\right)\right)$. Furthermore, $f_{e_{i}^{\prime} Y \mid W, e_{j}^{\prime} Y}\left(y_{i}\right)$ is bounded above by an integrable function $g\left(y_{i}\right)$ on $\mathbb{R}$ and bounded below by some constant $C>0$ in a neighborhood of $\left(w^{*}, k_{\alpha, j}\left(w^{*}\right)\right)$.
  \item The conditional quantile $k_{\gamma, e_{i}}\left(w^{*}, k_{\alpha, j}\left(w^{*}\right)\right)$ is partially differentiable with respect to both arguments at $\left(w^{*}, k\left(\alpha, e_{j} \mid w^{*}\right)\right)$. Similarly, the marginal quantile $k_{\alpha, j}\left(w^{*}\right)$ is partially differentiable with respect to any component of $(p, x)$ at $w^{*}$.
  \item The function $\phi(p, x, u)$ is approximately differentiable in all dimensions of $w_{1}:=(p, x) \in \mathbb{R}^{L}$ in the sense that there exist measurable functions $\Delta_{s}, s=1, \ldots, L$ satisfying
\begin{multline*} 
P\left[\left|\phi\left(w_{1 s}^{*}+\delta, w_{-1 s}^{*}, z^{*}, A\right)-\phi\left(w_{1 s}^{*}, w_{-1 s}^{*}, z^{*}, A\right)-\delta \Delta_{s}(A)\right|\right. \\ 
\geq \left.\delta \varepsilon \mid W=w^{*}, e_{j}^{\prime} Y=k_{\alpha, j}\left(w^{*}\right)\right]=o(\delta).
\end{multline*}
for $\delta \rightarrow 0$ and fixed $\varepsilon>0$. Analogously to Dette, Hoderlein $\delta$ Neumeyer (2016), we write $\partial_{w_{1 s}} \phi\left(w_{1}^{*}, z^{*}, a\right):=\Delta_{s}(a)$ and $\partial_{w_{1 s}} \phi:=\Delta_{s}(A)$ for all $s=1, \ldots, L$.
  \item The conditional distribution of $\left(e_{i}^{\prime} Y, \partial_{w_{1} s} \phi\right)$ given $\left(P, X, Z, e_{j}^{\prime} Y\right)$ is absolutely continuous with respect to Lebesgue measure for $\left(W, e_{j}^{\prime} Y\right)=\left(w^{*}, k_{\alpha, j}\left(w^{*}\right)\right)$ and all $i, j \in\{1, \ldots, L\}$. The conditional density $f_{Y_{i}, \partial_{w_{1 s}} \phi \mid W, e_{j}^{\prime} Y}$ satisfies

$$
f_{Y_{i}, \partial_{w_{1 s}} \phi \mid W=w^{*}, e_{j}^{\prime} Y=k_{\alpha, j}\left(w^{*}\right)}\left(y, y^{\prime}\right) \leq C g\left(y^{\prime}\right)
$$

for some constant $0<C<+\infty$ and a positive density function $g\left(y^{\prime}\right)$ with finite first moment on $\mathbb{R}$.
\end{enumerate}
\end{assumption}

\subsection{Proof of Lemma \ref{lem:hod_mam}}
\begin{proof} The proof is similar to the proof of Theorem 2.1 in \citet{Hoderlein2007}, except for several additional terms and computations. To reduce notational clutter, we assume that $W$ is univariate in analogy to \citet{Hoderlein2007} (this is without loss of generality since the argument applies to each component of $w$ separately, and the multivariate case follows by repeating the derivation for each partial derivative $\partial_{w_{1s}}$, $s = 1, \ldots, L$). By definition of the conditional quantile we have
\begin{align*}
0 &= P\!\left(e_i'Y \leq k_{\gamma_{i|j},i}(w^*+\delta, k_{\alpha_j,j}(w^*+\delta)) \;\middle|\; W = w^*+\delta,\, e_j'Y = k_{\alpha_j,j}(w^*+\delta)\right) \\
&\quad - P\!\left(e_i'Y \leq k_{\gamma_{i|j},i}(w^*, k_{\alpha_j,j}(w^*)) \;\middle|\; W = w^*,\, e_j'Y = k_{\alpha_j,j}(w^*)\right) \\
&= A_1 + A_2 + A_3,
\end{align*}
where
\begin{align*}
A_1 &= P\!\left(e_i'\phi(w^*+\delta, U) \leq k_{\gamma_{i|j},i}(w^*+\delta, k_{\alpha_j,j}(w^*+\delta)) \;\middle|\; W = w^*+\delta,\, e_j'Y = k_{\alpha_j,j}(w^*+\delta)\right) \\
&\quad - P\!\left(e_i'\phi(w^*+\delta, U) \leq k_{\gamma_{i|j},i}(w^*, k_{\alpha_j,j}(w^*)) \;\middle|\; W = w^*+\delta,\, e_j'Y = k_{\alpha_j,j}(w^*+\delta)\right), \\[6pt]
A_2 &= P\!\left(e_i'\phi(w^*+\delta, U) \leq k_{\gamma_{i|j},i}(w^*, k_{\alpha_j,j}(w^*)) \;\middle|\; W = w^*+\delta,\, e_j'Y = k_{\alpha_j,j}(w^*+\delta)\right) \\
&\quad - P\!\left(e_i'\phi(w^*+\delta, U) \leq k_{\gamma_{i|j},i}(w^*, k_{\alpha_j,j}(w^*)) \;\middle|\; W = w^*,\, e_j'Y = k_{\alpha_j,j}(w^*)\right), \\[6pt]
A_3 &= P\!\left(e_i'\phi(w^*+\delta, U) \leq k_{\gamma_{i|j},i}(w^*, k_{\alpha_j,j}(w^*)) \;\middle|\; W = w^*,\, e_j'Y = k_{\alpha_j,j}(w^*)\right) \\
&\quad - P\!\left(e_i'\phi(w^*, U) \leq k_{\gamma_{i|j},i}(w^*, k_{\alpha_j,j}(w^*)) \;\middle|\; W = w^*,\, e_j'Y = k_{\alpha_j,j}(w^*)\right).
\end{align*}

Consider each term step by step. We have
\begin{align*}
A_1 &= \int_{k_{\gamma_{i|j},i}(w^*, k_{\alpha_j,j}(w^*))}^{k_{\gamma_{i|j},i}(w^*+\delta, k_{\alpha_j,j}(w^*+\delta))} f_{Y_i|W=w^*+\delta, Y_j = k_{\alpha_j,j}(w^*+\delta)}(y_i)\,dy_i \\
&= \delta\Big[\partial_{1} k_{\gamma_{i|j},i}(w^*, k_{\alpha_j,j}(w^*)) + \partial_2 k_{\gamma_{i|j},i}(w^*, k_{\alpha_j,j}(w^*))\,\partial_w k_{\alpha_j,j}(w^*)\Big] \\
&\qquad \cdot f_{Y_i|W=w^*, Y_j = k_{\alpha_j,j}(w^*)}(k_{\gamma_{i|j},i}(w^*, k_{\alpha_j,j}(w^*))) + o(\delta),
\end{align*}
which follows from parts~1 and~3 of Assumption~\ref{ass:reg} and the multivariate chain rule for $\delta \to 0$.

For $A_2$:
\begin{align*}
A_2 &= \int_{-\infty}^{k_{\gamma_{i|j},i}(w^*, k_{\alpha_j,j}(w^*))} \Big[f_{Y_i|W=w^*+\delta, Y_j = k_{\alpha_j,j}(w^*+\delta)}(y_i) - f_{Y_i|W=w^*, Y_j = k_{\alpha_j,j}(w^*)}(y_i)\Big]\,dy_i \\
&= \delta\int_{-\infty}^{k_{\gamma_{i|j},i}(w^*, k_{\alpha_j,j}(w^*))} \Big[\partial_1 f_{Y_i|W=w^*, Y_j = k_{\alpha_j,j}(w^*)}(y_i) \\
&\qquad\qquad\qquad\qquad\qquad + \partial_2 f_{Y_i|W=w^*, Y_j = k_{\alpha_j,j}(w^*)}(y_i)\,\partial_w k_{\alpha_j,j}(w^*)\Big]\,dy_i + o(\delta) \\
&= \delta\Big[\partial_{w_{1s}} F_{Y_i|W=w^*, Y_j = k_{\alpha_j,j}(w^*)}(k_{\gamma_{i|j},i}(w^*, k_{\alpha_j,j}(w^*))) \\
&\qquad + \partial_w k_{\alpha_j,j}(w^*)\,\partial_2 F_{Y_i|W=w^*, Y_j = k_{\alpha_j,j}(w^*)}(k_{\gamma_{i|j},i}(w^*, k_{\alpha_j,j}(w^*)))\Big] + o(\delta),
\end{align*}
where the second line follows from parts~1, 2, and~3 of Assumption~\ref{ass:reg} in combination with the dominated convergence theorem and the multivariate chain rule for $\delta \to 0$. Note that the first term $\partial_{w_{1s}} F_{Y_i|W=w^*, Y_j = k_{\alpha_j,j}(w^*)}$ does \emph{not} vanish: while Assumption~\ref{ass:ind} ensures $A \perp\!\!\!\perp (P,X) \mid Z$, the conditional distribution $F_{Y_i|W, Y_j = y_j}$ conditions on the level set $\{a : e_j'\phi(w_1, z^*, a) = y_j\}$, which depends on $w_1$ through $\phi$. This distinguishes the bivariate setting from the univariate case in \citet{Hoderlein2007}, where no such conditioning is present and the analogous term is zero.

By the same argument as in \citet{Hoderlein2007}, under parts~1, 4, and~5 of Assumption~\ref{ass:reg} we obtain
\begin{align*}
A_3 &= -\delta\,E\!\left[\partial_w\phi \;\middle|\; W = w^*,\, e_i'Y = k_{\gamma_{i|j},i}(w^*, k_{\alpha_j,j}(w^*)),\, e_j'Y = k_{\alpha_j,j}(w^*)\right] \\
&\qquad \cdot f_{Y_i|W=w^*, Y_j = k_{\alpha_j,j}(w^*)}(k_{\gamma_{i|j},i}(w^*, k_{\alpha_j,j}(w^*))) + o(\delta).
\end{align*}

Rewriting and simplifying the equation $A_1 + A_2 + A_3 = 0$ gives the claim.
\end{proof}

\subsection{Proof of Theorem \ref{thm:symmetry}}
\begin{proof}
$\mathcal{S}$ is symmetric if and only if $e_i'\mathcal{S}\,e_j = e_j'\mathcal{S}\,e_i$ for all unit vectors $e_i, e_j$. Fix a point $(y_i^*, y_j^*)$ in the interior of the support of $(Y_i, Y_j)$ given $W = w^*$. Define $\alpha_i, \alpha_j, \gamma_{i|j}, \gamma_{j|i}$ as in the statement of the theorem. This point lies on the isoquant $K_{ij}(\beta \mid w^*)$ for $\beta := F_{Y_i, Y_j \mid W = w^*}(y_i^*, y_j^*)$.

If $\mathcal{S}$ is symmetric, then
\begin{equation}\label{eq:sym_cond}
E\!\left[e_i'\mathcal{S}\,e_j \;\middle|\; W = w^*, (Y_i, Y_j) = (y_i^*, y_j^*)\right] = E\!\left[e_j'\mathcal{S}\,e_i \;\middle|\; W = w^*, (Y_i, Y_j) = (y_i^*, y_j^*)\right].
\end{equation}

Although $K_{ij}(\beta \mid w^*)$ is set-valued, the disintegration theorem (the conditions of Theorem~1 in \citet{chang1997conditioning} are satisfied since our measure is Radon on $\mathbb{R}$) provides two unique decompositions of the conditioning at the point $(y_i^*, y_j^*)$:

\begin{enumerate}
\item[(i)] Condition first on $Y_j = y_j^* = k_{\alpha_j,j}(w^*)$, then on $Y_i = y_i^* = k_{\gamma_{i|j},i}(w^*, k_{\alpha_j,j}(w^*))$;
\item[(ii)] Condition first on $Y_i = y_i^* = k_{\alpha_i,i}(w^*)$, then on $Y_j = y_j^* = k_{\gamma_{j|i},j}(w^*, k_{\alpha_i,i}(w^*))$.
\end{enumerate}

By uniqueness of the disintegration, the $\sigma$-algebras generated by these two decompositions coincide at the point $(y_i^*, y_j^*)$, so that
\begin{align*}
&E\!\left[e_i'\mathcal{S}\,e_j \;\middle|\; W = w^*, Y_i = k_{\gamma_{i|j},i}(w^*, k_{\alpha_j,j}(w^*)),\, Y_j = k_{\alpha_j,j}(w^*)\right] \\
&\quad = E\!\left[e_j'\mathcal{S}\,e_i \;\middle|\; W = w^*, Y_j = k_{\gamma_{j|i},j}(w^*, k_{\alpha_i,i}(w^*)),\, Y_i = k_{\alpha_i,i}(w^*)\right].
\end{align*}

The proof now proceeds similarly to the one in \citet{DETTE2016129}, using Lemma~\ref{lem:hod_mam}. Consider the left-hand side only, as the analysis for the right-hand side is analogous (with the roles of $i,j$ and the indices $\gamma_{j|i}, \alpha_i$ exchanged). By definition of the Slutsky matrix it holds
\begin{align*}
&E\!\left[\mathcal{S} \;\middle|\; P = p^*, X = x^*, Z = z^*, e_i'Y = k_{\gamma_{i|j},i}(w^*, k_{\alpha_j,j}(w^*)),\, e_j'Y = k_{\alpha_j,j}(w^*)\right] \\
&= E\!\left[D_p\phi \;\middle|\; W = w^*, e_i'Y = k_{\gamma_{i|j},i}(w^*, k_{\alpha_j,j}(w^*)),\, e_j'Y = k_{\alpha_j,j}(w^*)\right] \\
&\quad + E\!\left[\partial_x\phi\,\phi' \;\middle|\; W = w^*, e_i'Y = k_{\gamma_{i|j},i}(w^*, k_{\alpha_j,j}(w^*)),\, e_j'Y = k_{\alpha_j,j}(w^*)\right] \\
&= B_1 + B_2.
\end{align*}

Consider each term separately. Start with
\[
e_i'B_1\,e_j = e_i' E\!\left[D_p\phi \;\middle|\; W = w^*, e_i'Y = k_{\gamma_{i|j},i}(w^*, k_{\alpha_j,j}(w^*)),\, e_j'Y = k_{\alpha_j,j}(w^*)\right] e_j.
\]
Then by the same reasoning as \citet{DETTE2016129} one can write
\begin{align*}
e_i'B_1\,e_j &= E\!\left[e_i'D_p\phi \;\middle|\; W = w^*, e_i'Y = k_{\gamma_{i|j},i}(w^*, k_{\alpha_j,j}(w^*)),\, e_j'Y = k_{\alpha_j,j}(w^*)\right] e_j \\
&= E\!\left[\nabla_p e_i'\phi \;\middle|\; W = w^*, e_i'Y = k_{\gamma_{i|j},i}(w^*, k_{\alpha_j,j}(w^*)),\, e_j'Y = k_{\alpha_j,j}(w^*)\right] e_j \\
&= \nabla_{p,1} k_{\gamma_{i|j},i}(w^*, k_{\alpha_j,j}(w^*))\,e_j \\
&\quad + \left[\partial_2 k_{\gamma_{i|j},i}(w^*, k_{\alpha_j,j}(w^*)) + \frac{\partial_2 F_{Y_i|W=w^*, Y_j = k_{\alpha_j,j}(w^*)}(k_{\gamma_{i|j},i}(w^*, k_{\alpha_j,j}(w^*)))}{f_{Y_i|W=w^*, Y_j = k_{\alpha_j,j}(w^*)}(k_{\gamma_{i|j},i}(w^*, k_{\alpha_j,j}(w^*)))}\right] \nabla_p k_{\alpha_j,j}(w^*)\,e_j \\
&\quad + \frac{\nabla_{p,1} F_{Y_i|W=w^*, Y_j = k_{\alpha_j,j}(w^*)}(k_{\gamma_{i|j},i}(w^*, k_{\alpha_j,j}(w^*)))}{f_{Y_i|W=w^*, Y_j = k_{\alpha_j,j}(w^*)}(k_{\gamma_{i|j},i}(w^*, k_{\alpha_j,j}(w^*)))} \cdot e_j,
\end{align*}
where the last equality follows from Lemma~\ref{lem:hod_mam}.

For the second term it holds
\begin{align*}
e_i'B_2\,e_j &= E\!\left[e_i'\partial_x\phi\,\phi'\,e_j \;\middle|\; P = p^*, X = x^*, Z = z^*, e_i'Y = k_{\gamma_{i|j},i}(w^*, k_{\alpha_j,j}(w^*)),\, e_j'Y = k_{\alpha_j,j}(w^*)\right] \\
&= E\!\left[\partial_x e_i'\phi \cdot \phi'\,e_j \;\middle|\; W = w^*, e_i'Y = k_{\gamma_{i|j},i}(w^*, k_{\alpha_j,j}(w^*)),\, e_j'Y = k_{\alpha_j,j}(w^*)\right] \\
&= E\!\left[\partial_x e_i'\phi \;\middle|\; W = w^*, e_i'Y = k_{\gamma_{i|j},i}(w^*, k_{\alpha_j,j}(w^*)),\, e_j'Y = k_{\alpha_j,j}(w^*)\right] k_{\alpha_j,j}(w^*) \\
&= \partial_{x,1} k_{\gamma_{i|j},i}(w^*, k_{\alpha_j,j}(w^*))\, k_{\alpha_j,j}(w^*) \\
&\quad + \left[\partial_2 k_{\gamma_{i|j},i}(w^*, k_{\alpha_j,j}(w^*)) + \frac{\partial_2 F_{Y_i|W=w^*, Y_j = k_{\alpha_j,j}(w^*)}(k_{\gamma_{i|j},i}(w^*, k_{\alpha_j,j}(w^*)))}{f_{Y_i|W=w^*, Y_j = k_{\alpha_j,j}(w^*)}(k_{\gamma_{i|j},i}(w^*, k_{\alpha_j,j}(w^*)))}\right] \partial_x k_{\alpha_j,j}(w^*)\, k_{\alpha_j,j}(w^*) \\
&\quad + \frac{\partial_{x,1} F_{Y_i|W=w^*, Y_j = k_{\alpha_j,j}(w^*)}(k_{\gamma_{i|j},i}(w^*, k_{\alpha_j,j}(w^*)))}{f_{Y_i|W=w^*, Y_j = k_{\alpha_j,j}(w^*)}(k_{\gamma_{i|j},i}(w^*, k_{\alpha_j,j}(w^*)))}\, k_{\alpha_j,j}(w^*),
\end{align*}
where the third line follows from the conditioning ($e_j'Y = \phi'\,e_j = k_{\alpha_j,j}(w^*)$ is in the conditioning set) and the last equality follows each time from Lemma~\ref{lem:hod_mam}. The same result holds when switching the roles of $i$ and $j$ (using decomposition~(ii) with indices $\gamma_{j|i}, \alpha_i$), and subtraction and rearranging gives the claim. 
\end{proof}

\end{document}